# SciJava Ops: An Improved Algorithms Framework for Fiji and Beyond

Gabriel J. Selzer[1], Curtis T. Rueden[1], Mark C. Hiner[1], Edward L. Evans III[1,2], David Kolb[3,4], Marcel Wiedenmann[3,4], Christian Birkhold[3,4], Tim-Oliver Buchholz[5,6], Stefan Helfrich[4], Brian Northan[7], Alison Walter[1,2,4], Johannes Schindelin[1,2,5,8], Tobias Pietzsch[5], Stephan Saalfeld[5,9], Michael R. Berthold[3,4], and Kevin W. Eliceiri[1,2]*.

[1] Center for Quantitative Cell Imaging, University of Wisconsin–Madison, Madison, Wisconsin, USA

[2] Morgridge Institute for Research, Madison, Wisconsin, USA
[3] University of Konstanz, Konstanz, Germany
[4] KNIME GmbH, Konstanz, Germany
[5] Max Planck Institute of Molecular Cell Biology and Genetics, Dresden, Germany
[6] Friedrich Miescher Institute for Biomedical Research, Basel, Switzerland
[7] True North Intelligent Algorithms
[8] Microsoft Corporation
[9] Janelia Research Campus, Howard Hughes Medical Institute, Ashburn, Virginia, USA

**\* Correspondence:**
Corresponding Author
eliceiri@wisc.edu

**Keywords: algorithms, image analysis, scientific imaging, imaging software, analysis workflows, Fiji, OpenCV, Python**

# 1 Abstract

Decades of iteration on scientific imaging hardware and software has yielded an explosion in the size, complexity, and heterogeneity of image datasets and the many different computational approaches employed to analyze them. The wide array of resulting imaging software tools and platforms present a challenge for researchers, data scientists, and engineers to understand and navigate: they must not only invest time becoming familiar with emerging computational tools, but must also decide how to integrate and deploy their own tools to foster adoption within the scientific community. Many scientific software platforms provide plugin mechanisms that simplify the integration, deployment, and execution of externally developed functionality. One of the most widely used platforms in the imaging space is Fiji, a popular open-source application for scientific image analysis. Fiji incorporates and builds on the ImageJ and ImageJ2 platforms, which provide a powerful plugin architecture used by thousands of plugins to solve a wide variety of problems. This capability is a major part of Fiji's success, and it has become a widely used biological image analysis tool and a target for new functionality. However, a plugin-based software architecture cannot unify disparate platforms operating on incompatible data structures; interoperability necessitates the creation of adaptation or "bridge" layers to translate data and invoke functionality. As a result, while platforms like Fiji enable a high degree of interconnectivity and extensibility, they were not fundamentally designed to integrate across the many data types, programming languages, and architectural differences of various software platforms. To help address this challenge, we present SciJava Ops, a foundational software library for expressing algorithms as plugins in a unified and extensible way. Continuing the evolution of Fiji's SciJava plugin mechanism, SciJava Ops enables users to harness algorithms from various software platforms within a central execution environment. In addition, SciJava Ops automatically adapts data into the most appropriate structure for each algorithm, allowing users to freely and transparently combine algorithms from otherwise incompatible tools. While SciJava Ops is initially distributed as a Fiji update site, the framework does not require Fiji, ImageJ, or ImageJ2, and would be suitable for integration with additional image analysis platforms.

## 2 Introduction

Scientific image analysis is, at its core, the act of applying algorithms to image data to explore a hypothesis. Image analysts need to extract knowledge from raw image data, the numerical signals recorded from an instrument. To do this, they need algorithms: mathematical routines to interrogate, transform, and distill the data into more refined forms; for example, segmenting a biological image into regions of interest (ROI) such as cells. To apply these algorithms at scale across large quantities of image data, they must be expressed as computer programs. Because building computational tools is a lengthy and painstaking process, software developers construct solutions on top of existing foundations whenever possible. But there are many diverging foundational layers from which to choose, including various machine architectures, operating systems, programming languages, and software libraries. Complicating matters, image datasets are stored in a variety of ways, such as on the cloud (1), on the GPU (2), or in memory; their values could be represented in numerous ways such as 8-bit integers (0-255) or floating point (a wide range but with some limits on numerical precision); and these values could be organized along rows, columns, strips, tiles, or in multidimensional blocks in one of many different file formats (3–5). Each choice imposes its own costs and benefits, and to allow computation within different settings a huge number of image analysis routines have naturally proliferated.

As the quantity and diversity of algorithm implementations has grown, there have been many efforts to organize them into cohesive software platforms to the benefit of users. Such platforms serve to unify access to compatible algorithms by operating on common data structures with defined conventions, making it easier to link them into workflows: sequences of operations where each algorithm is executed as a step in a larger analysis process. But each platform must be grounded in specific technical choices. Scientific communities tend to be partitioned by programming language as integration across language boundaries adds substantial complexity to software design. Some bioscience platforms build on the Java Virtual Machine (JVM), including ImageJ (6), Fiji (7), QuPath (8), and Icy (9). Many others build on the Python programming language and its *n*-dimensional NumPy array data structure (10) and SciPy extensions for scientific computing (11), including the scikit-image library (12), CellProfiler (13), and napari (14). Web-based tools are also increasingly prevalent: examples in the bioimaging space include ImJoy (15), the BioImage Model Zoo (16), Cytomine (17), and BIAFLOWS (18). In the field of medical imaging, C++-based software platforms have been developed and refined for decades, including the Insight Toolkit (ITK) (19) and Visualization Toolkit (VTK) (20).

Some platforms strive to integrate functionality across paradigms—e.g.: Jupyter Notebook (21) offers a visual browser-based application for developing and disseminating workflows written in a variety of languages including Python, Java, and others; KNIME Analytics Platform (22) is an extensible tool for data science including image processing (23) which combines a JVM-based desktop application with web-based server components and support for executing Python programs as workflow nodes; and OMERO (24) is a web-based platform for managing microscopy image data, accessible from Java, Python, C++, and MATLAB programs as well as via its OMERO.insight desktop application. Even with such integrations in place, however, most platforms are not naturally mutually compatible: e.g., an "image" in ITK is not the same thing technically as an ImageJ image or a NumPy array. This heterogeneity of image data structures can be frustrating for users, who often struggle to combine multiple tools, leading to the creation of numerous additional integration mechanisms to link otherwise incompatible platforms (25–28). Fortunately, Python has emerged as a common integration point for many scientific software platforms across these languages, with research software engineers putting in the extra effort required to make tools work from Python programs in addition to their original target platforms. The rise of deep-learning-based image analysis in the latter half of the 2010s (29–31) further cemented Python as effective common ground to unify algorithm access. In the case of ImageJ and Fiji, for example, we created the PyImageJ library (32) and napari-imagej plugin (33) to facilitate easier access to ImageJ and other JVM-based routines from Python programs. ITK similarly has developed Python bindings (34), as well as better integration into web technologies (35), to make its algorithms more accessible.

Nonetheless, users typically still convert their data explicitly and manage the minutiae of every included library, with additional effort required to combine each new technology, such that the conversion of images between tools and types is often complex, nuanced, and expensive. Furthermore, it remains a substantial challenge even knowing what algorithms are available in each of the many software platforms for image analysis, much less combining them into a common scientific workflow to effectively extract knowledge. The ImageJ Ops library, part of the ImageJ2 project aiming to expand ImageJ's utility (36), introduced a more unified algorithm development and discovery framework for ImageJ-based software, building on the powerful ImgLib2 library (37) to support *n*-dimensional image data regardless of how it is represented internally. By implementing algorithms as structured plugins with input and output parameters constrained to specific types, it became possible for users to perform declarative image processing, where users decide which algorithms to apply, while the ImageJ Ops infrastructure selects the best implementations for particular input data. The SciJava Ops project described in this manuscript and depicted in Figure 1 is an evolution of that effort which improves on the ImageJ Ops design while taking necessary steps to decouple from ImageJ and establish connections outside the JVM.

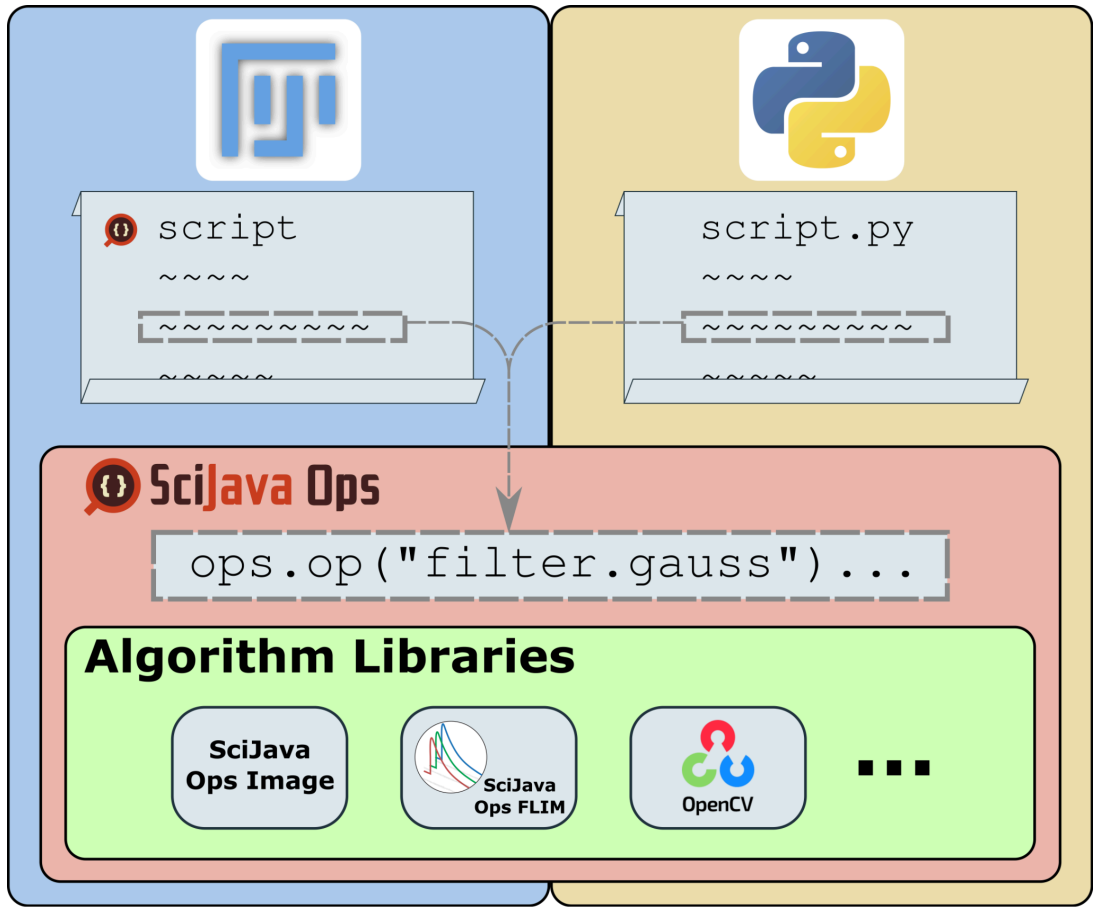

**Figure 1:** SciJava Ops provides access to a diverse suite of algorithms through a uniform syntax which allows the framework to match the most appropriate algorithm to each user request and allows users to freely combine otherwise incompatible algorithms into a streamlined workflow. All SciJava Ops algorithms are usable from any of Fiji's supported script languages (e.g. Groovy or Jython), as well as from Python scripts.

## 3 Method

### 3.1 Design Principles

The primary aim of SciJava Ops is to provide unified access to scientific algorithms on the Java Virtual Machine, while establishing a language-independent design for structured algorithms:

1. An **Op** is a plugin defining an algorithm implementation. It has a **name** indicating its purpose, and a variable number of input and output **parameters**, each defined by a language-specific **type**. For example, a function that adds two integers, `a` and `b`, producing the result as a new integer, might be represented by the Op `math.add(a:int, b:int)→int`. It is expected that Ops are **deterministic** and **reproducible**, yielding the same result when given the same **arguments** (parameter values). Ops adhere to **functional interfaces**: structural patterns that define common behavior.

- A **function** Op accepts immutable (unchanging) **input** values and produces a new **output** value. E.g.: `filter.gauss(input:image, sigma:float)→image` accepts an existing image as input, along with a real-valued sigma, and performs a Gaussian blur operation, producing a transformed image as new data, occupying additional computer memory, while leaving the original input image unchanged.
- A **computer** Op accepts immutable (unchanging) input values and writes its output into a pre-allocated **container**. E.g.: `filter.gauss(input:image, sigma:float, output:image)` accepts an existing image as input, along with a real-valued sigma, and performs a Gaussian blur operation, overwriting the contents of the container image (named output) with the transformed data. In this way, no new computer memory needs to be allocated for the computation.
- An **inplace** Op accepts input values, one of which is **mutable** and overwritten with the transformed data. E.g.: `filter.gauss(input:image, sigma:float)` performs a Gaussian blur operation on the input image, overwriting its contents with the computation result, thus destroying the original data values, but saving space by avoiding allocation of additional computer memory.

Each of these three patterns have advantages and disadvantages, and as each is frequently seen (though typically less formally) in existing libraries of algorithms, all three are necessary to accommodate the broadest range of potential Ops.

2. Ops are declared via metadata descriptor files, enabling all Ops to be cataloged in a unified way, independent of language and implementation. Each entry in one of these descriptor files references an Op written for a particular target platform, such as a Python function or Java method. While these files alone are not enough to actually execute any particular Op, they are sufficient to reason about which Ops would be available. Descriptor files also enable **zero-code wrapping**: algorithmic code from existing libraries can be made available within SciJava Ops *without being written for* SciJava Ops.

3. All available Ops are collected into an Op **environment** which provides access to Ops functionality. Via this environment, Ops are **requested** by name and parameter types. The intent of the design is to allow users to specify the algorithm they want to run, without mandating which specific implementation will be used. This approach leaves room for the best implementation to be chosen each time an algorithm is requested, enabling existing Ops-based workflows to automatically improve over time as better Ops are added to the environment. For example, an image processing script that calls `filter.gauss` may become faster when a GPU-based Gaussian filter Op becomes available.

4. Op execution requests are powered by the Ops **engine**: a matching system that chooses the most appropriate available Op from the environment. The engine takes into account not only the number and types of parameters, but also the type of functional interface (function, computer, or inplace). In case no Op exists with strictly matching parameter types, the engine can implicitly convert parameters into other types that *do* match an existing Op. For example, an Op written to convolve an ITK image could be used even when the user requests to convolve an ImgLib2 image, without the user needing to be cognizant of ITK data structures, as long as the environment has `engine.convert` Ops available capable of converting between those two types of images.

5. Ops are atomic units that can build on each other. For example, a Difference of Gaussians (`dog`) Op:

   `filter.dog(input:image, sigma1:float, sigma2:float)→image`

might leverage helper Ops as **Op dependencies** to execute the formula:

   `filter.gauss(image, sigma1) - filter.gauss(image, sigma2)`

where the minus (-) operator is executed using a `math.sub(a, b)` Op, as shown in (C) of Figure 2 below. Such dependencies are also decided by the engine, rather than linking to specific Ops. For example, if the more performant GPU-based `filter.gauss` Op described in (3) above is added to the environment, then the `filter.dog` Op described here would also benefit automatically.

## 3.2 Architecture

The Ops design elements above are independent of any programming language—but ultimately the design must be implemented as a software library built on a concrete foundation. We chose the Java Virtual Machine to implement the Ops design for several reasons: 1) the widespread adoption and ease of use of the Fiji platform, written in Java, makes it an ideal initial deployment target across computing platforms (38); 2) the ImageJ Ops library served as a launching point to improve on the existing Ops design while providing a large number of image processing Ops as a foundation; and 3) ImageJ and Fiji functionality are already accessible from Python via the PyImageJ project, offering an immediate way for Ops to be used from Python as well, and later to wrap existing Python libraries into Ops.

SciJava Ops is partitioned into several different component libraries. The following sections describe the application programming interface (API) layers of SciJava Ops in increasing technical concern. We start with basic querying and use of Ops, move to development of new Ops, and end with technical overviews of the mechanisms powering Ops from user and developer perspectives.

3.2.1 Requesting Ops

The most common entry point into SciJava Ops is the `scijava-ops-api` component, which offers a way to request Ops declaratively by name and parameter types, using the **Op builder** mechanism of the Op environment class. The Op builder pattern partitions the request into a granular set of function calls, as illustrated in (A) of Figure 2; completed Op builder requests are then provided to internal SciJava Ops components to attempt to match an algorithm implementation.

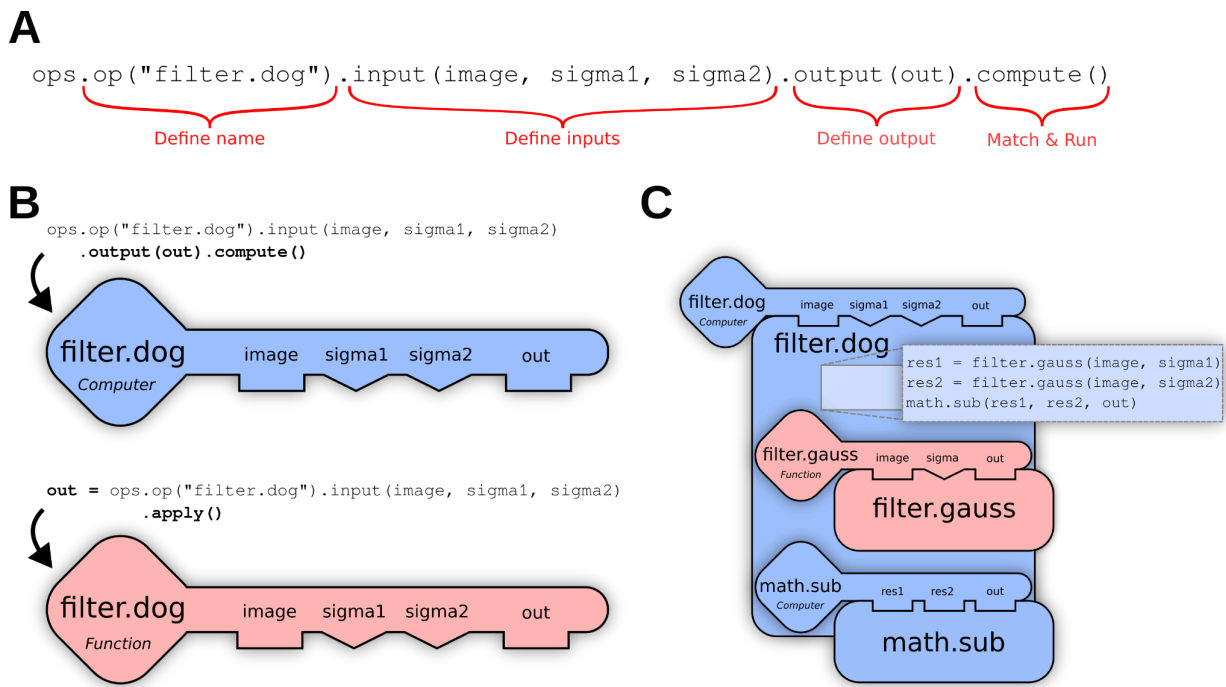

**Figure 2: (A)** Using the Op builder pattern, the user requests a `filter.dog` (Difference of Gaussians) Op accepting an ordered list of three input parameters—an image and two sigma values—as well as a suitable container into which the output of the computation will be stored. The terminating method call `compute()` triggers the Ops engine to search for a computer Op plugin capable of fulfilling the request. **(B)** Alternately, if the terminating method call is `apply()` instead of `compute()`, an output container does not need to be specified, and the Ops engine will search for a function Op, which generates and returns the result. **(C)** The matching `filter.dog` Op declares dependencies on the helper Ops

> `filter.gauss` and `math.sub` as described above, which
> are then recursively matched, after which the `filter.dog`
> Op is actually executed to perform the computation.

Several variations of the above pattern are possible, depending on whether the user wants to match and execute the Op immediately, or receive a reference to the matched Op that can be invoked repeatedly as desired. For example, if the query is terminated with `.function()` rather than `.apply()`, the matching `math.add` Op itself will be returned as a `BiFunction` object, Java's functional interface for binary functions. Other variations are relevant in case the user wants to match as an inplace or computer Op, rather than a function Op.

To assist the user with knowing which Ops are available to call, the Op environment also provides `.help()` and `.helpVerbose()` methods to obtain descriptions of all Ops matching a particular query. For example, `ops.help("math")` would list all Ops in the `math` category (e.g., `math.add` and `math.sub`). On the other hand, `ops.help("math.add")` would list all known implementations of the `add` Op in the `math` category, while `ops.helpVerbose("math.add")` would provide the same list but with detailed usage documentation. These help methods can be called at any stage of the builder, listing potential Op matches according to current criteria.

### 3.2.1.1 Type Descriptions

While all image analysis software ultimately operates on images, the naming of image data structures between software packages is not consistent; for example, images are represented using ImgLib2 `RandomAccessibleInterval` structures within Fiji, as `ndarray` objects within NumPy-based projects, and as `Mat` objects within OpenCV (39). Transparent integration and interoperability between these platforms requires not only strategies for conversion between these data structures, as described above, but also abstraction from platform-specific nomenclature.

Within SciJava Ops, the set of `engine.describe` Ops map each type to a common, simplified name; for images, we might use `engine.describe` Ops to describe `RandomAccessibleInterval`, `ndarray`, and `Mat` all as "image". Then, when the user asks for information about a given Op, each parameter is replaced with its description, resulting in text without any terminology specific to particular software packages.

### 3.2.2 Defining Ops

A critical design goal of SciJava Ops is to make it as convenient as possible to bring new Ops into the system. To extend the collection of Ops available for use, developers can implement new algorithms, as well as introduce existing libraries into the framework, in ways that require minimal to zero additional changes to source code or project dependencies.

For developers wishing to declare their routines as Ops directly in their implementation code, the `scijava-ops-indexer` component provides a simple way to do so using Javadoc: a special comment describing the purpose and behavior of a piece of Java code. A Java class, method, or field tagged with an `@implNote` line will trigger automatic production of an Ops metadata descriptor file in YAML format, thus making the Ops engine aware of the routines and their structure. Figure 3 depicts a static method that uses standard syntax and data structures, but includes an `@implNote` line in its Javadoc which registers the method as a `copy.array` computer Op:

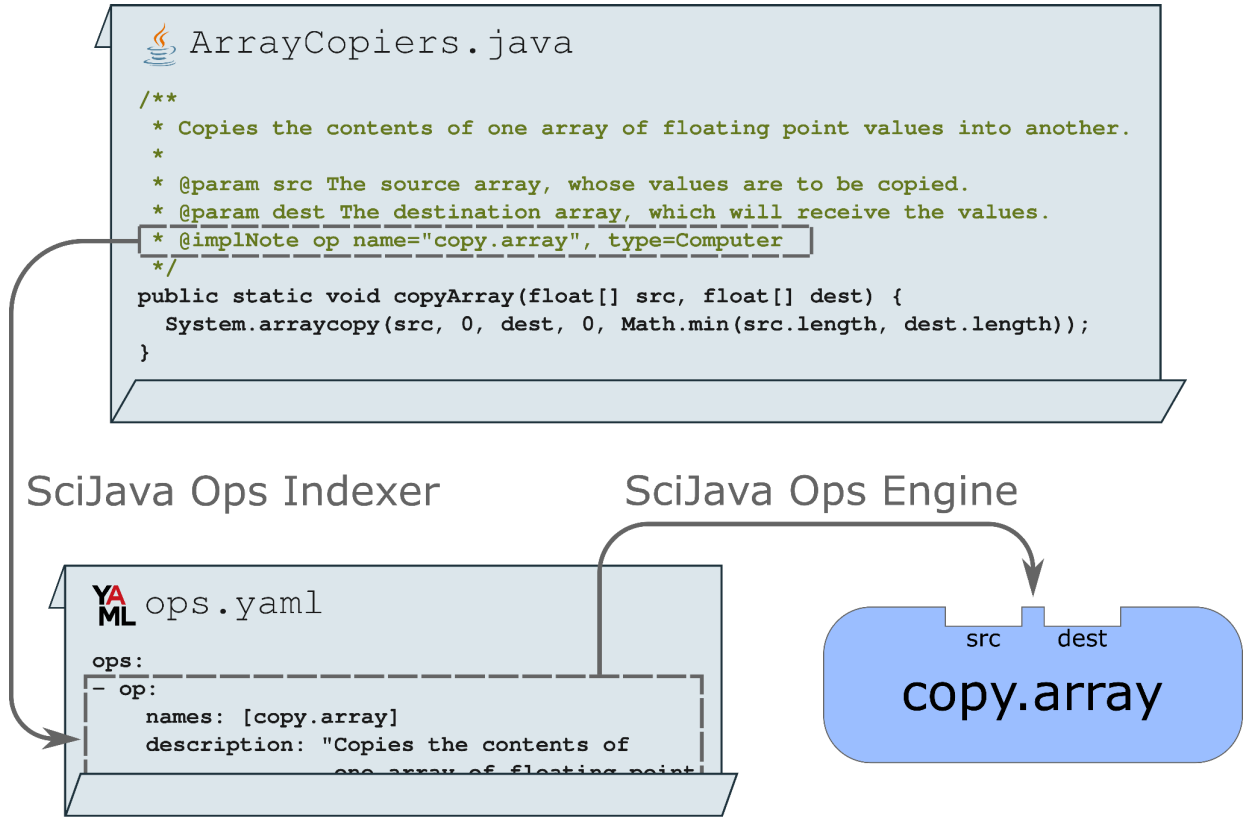

**Figure 3:** The SciJava Ops Indexer converts code blocks annotated with `@implNote` tags in their Javadoc into a YAML descriptor, containing metadata including the Op's name, description, and parameter types. At runtime, this YAML is parsed by the SciJava Ops Engine to create an Op.

Additional arguments, including the Op's name and priority, can be appended to this tag. It is important to note that this is a lightweight mechanism, allowing libraries to declare algorithms as Ops without explicitly requiring the Ops framework to be present to build or run their code outside of the Ops paradigm.

The Javadoc-based mechanism for defining Ops is convenient, but only usable by algorithm libraries written in Java, and requires modification to the source code's Javadoc sections. More generally—and especially in cases where such modifications are undesirable or impossible—the needed Ops YAML metadata descriptor files can be produced in other ways, making the Ops framework aware of code that was not originally written with Ops in mind. For example, third party libraries such as OpenCV can be made accessible as Ops without modifying the OpenCV code (see "OpenCV" in the "Example Use Cases" section below).

3.2.2.1 Framework Features

Libraries that choose to depend explicitly on SciJava components have access to additional useful mechanisms to facilitate the creation of better Ops. These mechanisms are used throughout the `scijava-ops-image` library to help them behave in a consistent and coordinated fashion.

- **`scijava-ops-spi`:** For Java developers who like strong type safety, the `scijava-ops-spi` component library enables the declaration of Ops using Java annotations, as well as the use of Op dependencies for reusing Ops as building blocks as described in the "Design Principles" section above. When an Op with @OpDependency annotations is matched, its dependencies are then matched recursively, and through the matching process the Op obtains the same benefits that SciJava Ops offers its users. This can make complicated Ops significantly shorter and enable other developers to override the behavior of dependencies for improved performance.
- **`scijava-progress`**: During the execution of its algorithm, an Op may use the `scijava-progress` API to issue status reports regarding its actions, which programs like Fiji can use to render user interface elements such as progress bars that keep the user informed of the Op's progress. For example, an iterative deconvolution Op might issue an update after each iteration is applied.
- **`scijava-concurrent`**: It is often advantageous to perform expensive calculations on multiple threads at once. The `scijava-concurrent` API provides a unified and convenient way for Ops to coordinate the sharing of compute resources with one another without complicating Op definitions with additional threading-related parameters. By using this API, multiple Ops that execute simultaneously are each assigned appropriate compute resources, e.g. two Ops using two processors each from a total of four available.

3.2.3 Matching Ops

The Ops engine includes several layers known as **matching routines**, which provide powerful and flexible fulfillment of Op requests in ways beyond only the base implementing algorithm. A request to add two images element-wise can be fulfilled by an Op that adds two elements, which is "lifted" or mapped to the entire image. A request to convolve MATLAB matrices can be fulfilled by an Op that convolves ImgLib2 images, as long as the engine has access to engine.convert Ops that translate between those two image types. A request for a function Op to perform a Gaussian filter on an image can be fulfilled by an computer Op as long as there is an engine.create Op to produce the computer's pre-allocated container output. In the following sections we will give more detail on these and related mechanisms.

3.2.3.1 Direct Op Matching

This matching routine seeks an Op that can run directly on user inputs. Searching through all Ops with a name matching the user request, the default routine asserts that the input argument types conform to the required parameter types of the Op. While allowing declarative access, direct matching does not perform any more sophisticated transformations to the arguments or to the underlying Op.

3.2.3.2 Op Adaptation

Further matching routines focus on transforming available Ops to produce a match: for example, calling an Op as if it implemented a different functional type. We define this as **Op adaptation**, enabled by special engine.adapt Ops to perform conversions as needed. Using adaptation, users can indirectly execute a computer Op as a function Op or invoke element-wise Ops across an entire list or image, embedding these transformations system-wide instead of requiring modifications per algorithm. A graphical example of Op adaptation is shown in Figure 4, and practical examples can be seen in Section 4.2, where Op adaptation is used to execute pixel-wise and neighborhood-wise Ops across entire images as well as to create pre-allocated output buffers.

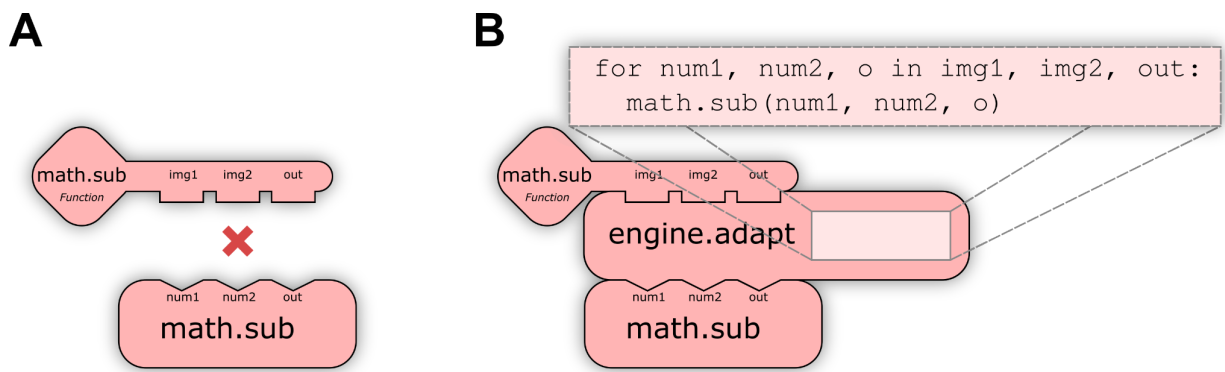

**Figure 4**: Op adaptation transforms an Op into a new Op with different structures. **(A)** The user requests an element-wise algebraic subtraction of two images (matrices of elements), but the only existing Op operates instead on individual elements, preventing a match. **(B)** Op adaptation can remedy such situations, by inserting an adaptation that iteratively executes the Op across all elements of the inputs.

3.2.3.3 Parameter Conversion

While Op adaptation alters the Op structure, a third matching routine called **parameter conversion** facilitates the conversion of individual inputs and outputs. The Ops engine's matching mechanism is fundamentally built around knowing the parameter types of each Op;

with this knowledge, multiple ops of the same name can exist that operate on different data types, such that the best Op for the job can be chosen in different scenarios. But with parameter conversion we can also take it a step further: rather than restricting Op requests to types aligned with each implementation—requiring users to exercise prior knowledge of the implementation details of available Ops—the engine can convert input arguments on the fly between types, such that a broader range of available Ops can be matched than would otherwise be possible. For example, the logic of convolving an ImgLib2 image with an OpenCV image is not truly different from convolving two ImgLib2 images: the application of a single ImgLib2 convolve Op on images of either or both image types represents the core SciJava Ops principle of interoperability.

Parameter conversion utilizes engine.convert Ops to convert a user's arguments to the Op implementation's required types, and similarly converts Op outputs to the user's requested output type. An example pathway is shown in Figure 5. As a result, the parameter conversion matching routine enables "fuzzy" matching, allowing users to execute Ops on a wider range of parameter types, removing the need for manual conversion beforehand or knowledge of exact Op implementations. A practical example of parameter conversion is described in Section 4.1, where it allows seamless execution of OpenCV algorithms using ImgLib2 data structures.

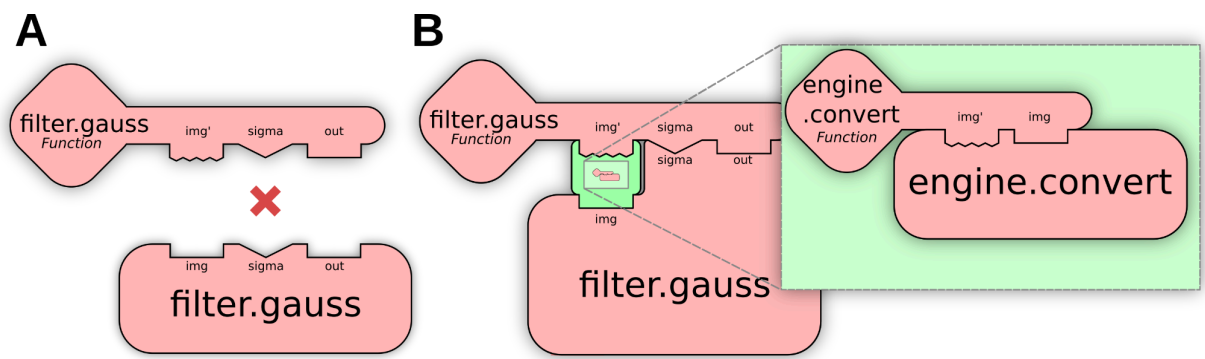

**Figure 5**: Parameter conversion allows "fuzzy matching" by transforming individual Op parameters. **(A)** The user requests a Gaussian Blur of an image, but the only existing Op is incompatible with the user's input image, preventing a match. **(B)** Parameter conversion enables a match by incorporating an engine.convert Op into the result, transforming the parameter before normal Op execution.

### 3.2.3.4. Optional Parameters

Optional function parameters are a convenient feature of many popular programming languages. For example, a rescale2D(image, width, height) function for resizing a 2-dimensional image plane might mark the height parameter as optional, so that rescale2D can be invoked with only image and width arguments, calculating the target height in a way that maintains the image's aspect ratio. SciJava Ops allows the use of optional parameters in an analogous fashion by creating "reduced" versions of the Op that omit parameters marked as optional one by one from right to left, in line with how most programming languages support this feature. For instance, one of the Fast Fourier Transform (FFT) Ops in SciJava Ops Image has the form `filter.fft(input, fftType, borderSize, fast) → output`, where the *borderSize* and *fast* parameters are both optional and each making a reasonable assumption if left off; this results in additional "reduced" Ops `filter.fft(input, fftType, borderSize) → output` and `filter.fft(input, fftType) → output` also being available in the Op environment. Optional parameters enable Ops to provide reasonable default behaviors while matching against more concise Op requests.

### 3.2.4 SciJava Ops Image

The `scijava-ops-image` component provides a large collection of image processing algorithms built on SciJava Ops, including functionality such as image filtering, thresholding, binary and grayscale image morphology operators, image segmentation and labeling, 2D and 3D object geometry, shape descriptors and measurements, numerical and image arithmetic, fast Fourier transformations, image convolution and deconvolution, spatial transformations, and colocalization. This library, which represents an evolution of the ImageJ Ops collection, is not intended to be exhaustive but instead foundational, offering Ops built on the ImgLib2 image library; other Ops built on other image libraries are equally possible.

## 4 Results

All components of the SciJava Ops framework are implemented in Java and deployed as libraries to the Maven Central repository, making them freely available for other software to build upon. SciJava Ops is also available in Fiji via the SciJava Ops update site, which provides access to and integration with its large and diverse plugin ecosystem (40–47). Detailed installation and usage instructions, as well as how-to guides and examples, can be found on the project website at https://ops.scijava.org/.

Through the architecture described above, SciJava Ops provides a framework for declarative, extensible, reusable algorithms, with a focus on image analysis. Diverse algorithms may be integrated seamlessly without depending upon them being written in a particular language, utilizing any particular data structures, or depending on specific features of SciJava Ops. The following use cases illustrate some of the possibilities already realized through the use of this framework.

### 4.1 OpenCV

The `scijava-ops-image` component provides a foundation for image analysis, but much like ImageJ and Fiji, the strength of SciJava Ops lies in its potential for extensibility. However, where ImageJ has provided a template for extension within its ecosystem, the vision for Ops is a grand unification of image processing libraries with integration layers built on top as needed, instead of requiring adherence to a shared implementation contract. In practice this has been accomplished through the use of YAML to identify Ops algorithms: the YAML serves as a separate deliverable which can be defined independently of the implementation library it references; as long as both are accessible to the SciJava Ops framework, the defined Ops will be available for use. To provide a practical example of this potential we developed the `scijava-ops-opencv` component.

OpenCV is a massive open source library for computer vision. Although OpenCV is written in C++, it has bindings for other languages—we targeted the bytedeco/javacv (https://github.com/bytedeco/javacv) Java wrappers for convenience. The bare minimum for inclusion into SciJava Ops was a Maven dependency on the bytedeco libraries and the YAML configuration file mapping the methods therein to Ops. There are over 2500 algorithms defined in OpenCV so we chose to focus our initial mappings to the Image Processing and Photo domains. Creating the YAML configuration file could be done manually, but this would be labor-intensive and tedious. Instead, we created a simple shorthand parsing program allowing the specification of OpenCV classes, methods, and essential metadata (e.g. the functional type) which then automatically identified cases such as overloaded methods and output appropriate Op YAML definitions. Thus with a cursory look at the OpenCV codebase, we were able to expose several hundred new methods into SciJava Ops.

However, true integration and interoperability also requires an exchange of data. To that end, we leveraged the prior development effort of ImageJ-OpenCV (https://github.com/imagej/imagej-opencv) which defines translation between the `Mat` data structure of OpenCV and `RandomAccessibleIntervals` of ImgLib2. Similar reuse of our shorthand configuration allowed the inclusion of ImageJ-OpenCV's translation methods into the SciJava Ops framework. One final element was required by SciJava Ops to take advantage of its adaptation and conversion mechanisms: the knowledge of how to copy between `Mat` data structures. Mats already have `copyTo` methods available in their API, so this simply required defining minimal copy Ops of our own that call this method. With these components in place, we could now use `Mats` and `RandomAccessibleIntervals` (and extensions thereof) interchangeably. Figure 6 provides a sense of how these software pieces all fit together when actually using an OpenCV Op.

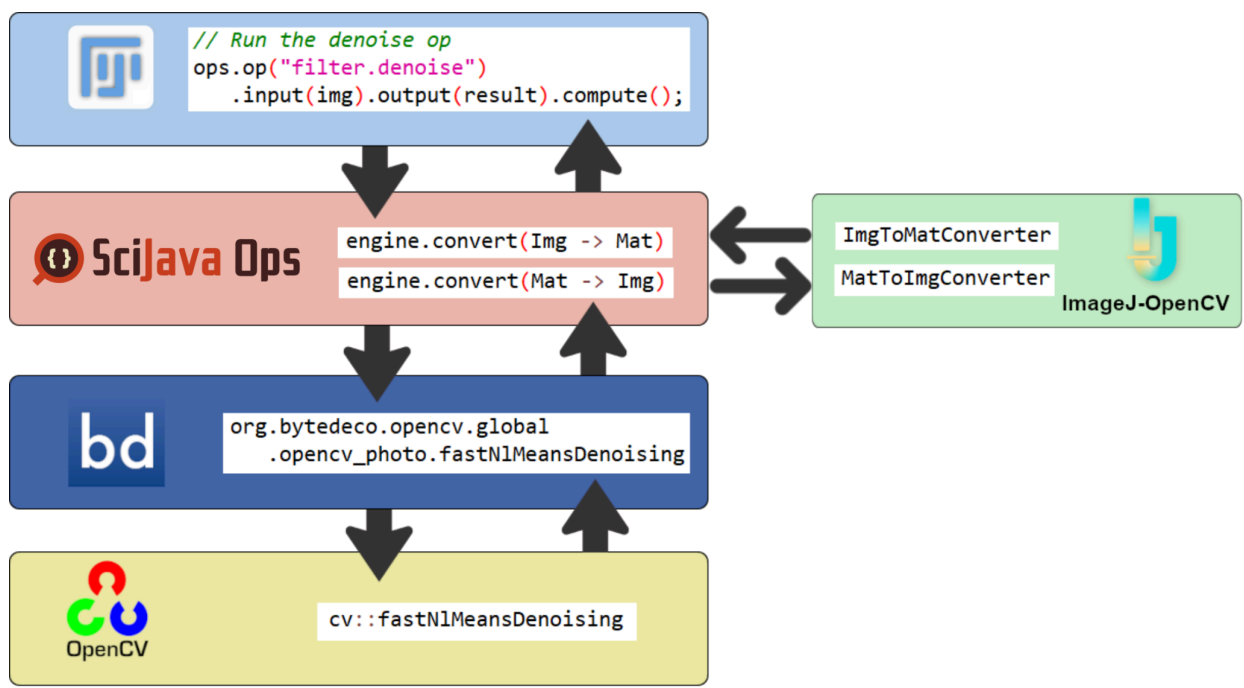

**Figure 6**: A high-level view of the flow of data through relevant components in calling an OpenCV Op. Starting from the top, when an Op is called, e.g. from a script in Fiji, SciJava Ops finds the appropriate function and performs necessary parameter conversion of data types. Control passes to the registered Bytedeco Java wrapper function, which itself makes the native C++ call to OpenCV. When data processing is complete our image returns up the layers, passing through another conversion back to the operative data type (ImgLib2's Img in this case).

There are several key observations highlighted in this figure. First, the OpenCV Non-Local Means denoise algorithm is being called in a general way through Ops: `filter.denoise` with the appropriate parameters. It is not identifiably tied to OpenCV. Second, we are providing the image data to the denoise algorithm as an ImgLib2 data structure despite the fact that the underlying OpenCV implementation operates on native `cv::Mat` instances. This conversion is happening automatically, transparent to the calling user. Perhaps most importantly, we can see the keystone role of SciJava Ops in unifying disparate software efforts for the benefit of connecting researchers with increased algorithm diversity.

## 4.2 Python (scyjava)

While we have made SciJava Ops accessible within Fiji via an ImageJ update site, there is nothing ImageJ- nor Fiji-specific about its implementation. Indeed the vision for SciJava Ops has always been to unite algorithm libraries across analysis tools and programming languages.

Python integration is of particular interest due to its prevalence in the scientific community and forefront position in the advancement of deep learning. Although the core of SciJava Ops is, naturally, written in Java, we can take advantage of the low-level scyjava and jgo libraries—built to facilitate PyImageJ—to gain immediate access from a Python entry point. Notably, SciJava Ops usage within a Python environment is identical to usage upon the JVM, allowing users to exercise familiarity with SciJava Ops within new workflow environments; once a user understands how to use SciJava Ops in one environment, they will be able to execute Ops in every environment.

In Figure 7 we see this process in-depth, starting from a minimal Python environment. As all SciJava Ops libraries are deployed to a public Maven repository, the scyjava configuration mechanism can succinctly be instructed to gather all necessary Java code and make it available in Python. At that point we can freely call Ops (or other Java code) as needed. For the sake of example we show Ops in use to perform basic image preprocessing, segmentation and measurement. While these particular Ops were written to operate on ImgLib2 data structures, the logic for converting from Pythonic data types (numpy) is already present in PyImageJ; registration of those methods as Ops would allow for seamless conversion and inter-use. Note that this use of scyjava would be appropriate from an interpreter or running as a script, but similarly enables SciJava Ops to be called from other Python tools, e.g. napari, as was done for the napari-imagej plugin.

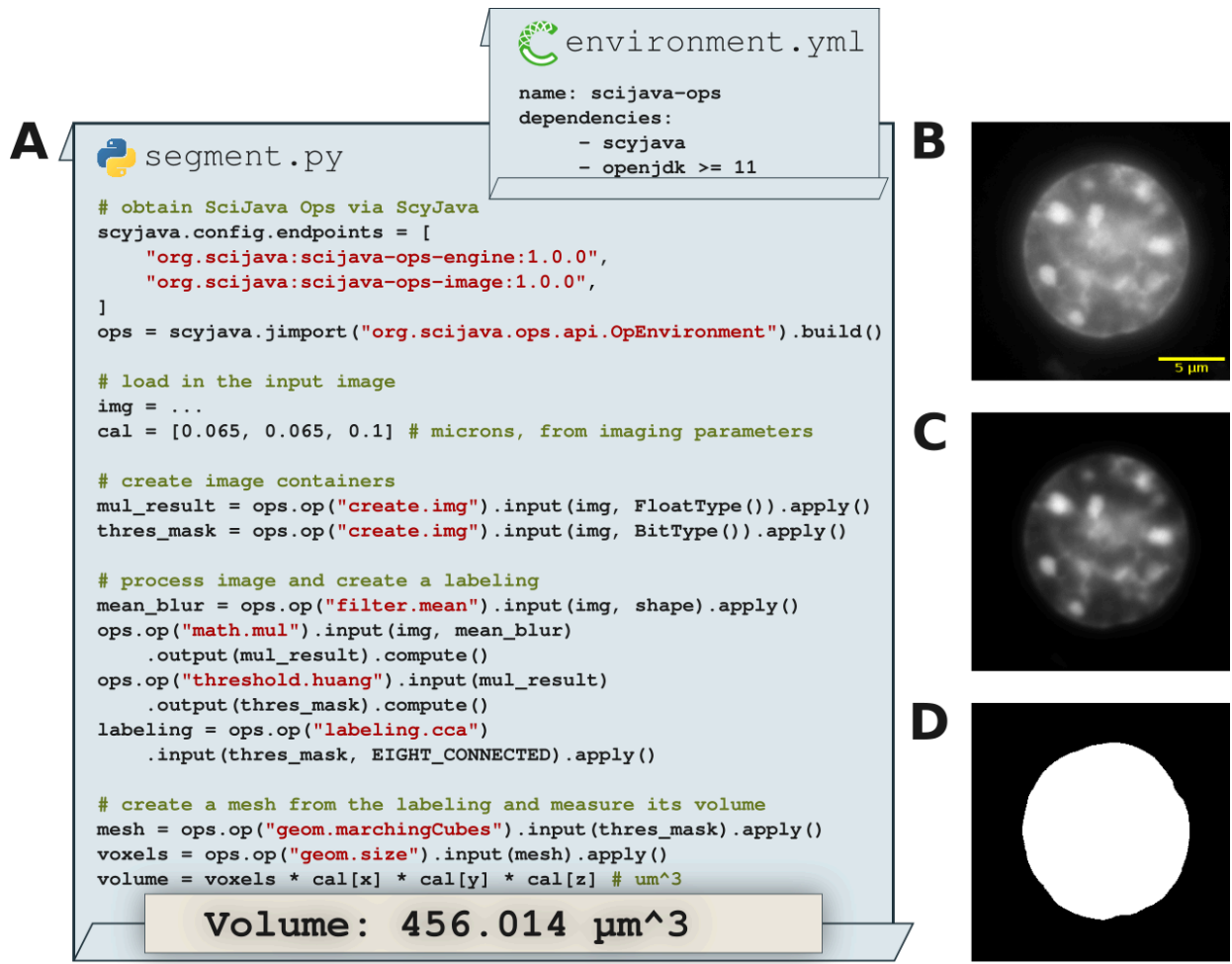

**Figure 7**: Image processing and volumetric 3D measurements in Python with SciJava Ops.

**(A)** Excerpts from sample Python code calling SciJava Ops via the scyjava library to image process input data, create a 3D mesh and measure its volume. **(B)** A slice of the input 3D (X, Y, Z) data of a 3T3 nucleus stained with DAPI. **(C)** A slice of the image processed result where an improved signal-to-noise ratio is achieved by multiplying the mean filtered image with the input using SciJava Ops. **(D)** Segmentation result of the image processed input data used to create the mesh.

## 4.3 Fluorescence Lifetime Image Analysis

Fluorescence Lifetime imaging microscopy (FLIM) utilizes the excited state decay rate of fluorophores to derive additional information about the surrounding microenvironment. Open, accessible tooling for FLIM analysis in Fiji is provided by FLIMJ (48), an ImageJ plugin wrapping the C++ library FLIMLib for use with ImgLib2 images. Included within FLIMJ is a set of algorithms written for the ImageJ Ops library, which we have translated into Ops for the SciJava Ops framework. The resulting Ops, contained within the supplementary SciJava Ops FLIM component, provide fluorescence decay rate fitting to one of many exponential-decay models, pseudo coloring of fit results, and subsequent post-processing. An example of SciJava Ops FLIM usage, performing a Levenburg-Marquardt algorithm (LMA) fit (49) and pseudo coloring the fluorescence lifetime map, is shown in Figure 8. Notably, SciJava Ops FLIM utilizes SciJava Ops' developer features to increase code reuse and extensibility within declared Ops. Whenever possible, FLIM Ops use Op dependencies to save lines of code, decreasing future maintenance and enabling future performance optimizations. They additionally utilize existing, shared data structures for seamless integration with both SciJava Ops Image and other Fiji functionality.

Figure 8 illustrates the transparent interoperability of SciJava Ops FLIM and existing Fiji components. Using the Bio-Formats plugin for Fiji (24), we load the image data into Fiji along with metadata required for FLIM analysis. Using ImageJ's selection tools, we can draw any number of ImageJ ROIs, which are then converted using Fiji to the ImgLib2 data structures used by SciJava Ops FLIM. In this case, the flexibility provided by these tools allows users to restrict computation to subregions of the data, avoiding the penalties of computation across undesired or uninteresting areas. With this script, users can observe fluorescence and the fluorescence lifetime across the region of interest, compute parameters of interest and could additionally perform segmentation, labeling or other post-processing routines.

## A

```groovy
// flim.groovy
#@ OpEnvironment ops
#@ Img img
// FitParams are inputs to FLIM fitting Ops
param = new FitParams(...)
param.transMap = img
// Fit curves
lma = ops.op("flim.fitLMA").input(param).apply()
// Finally, generate a pseudocolored result
pseudocolored = ops.op("flim.pseudocolor").input(lma).apply()
```

## B

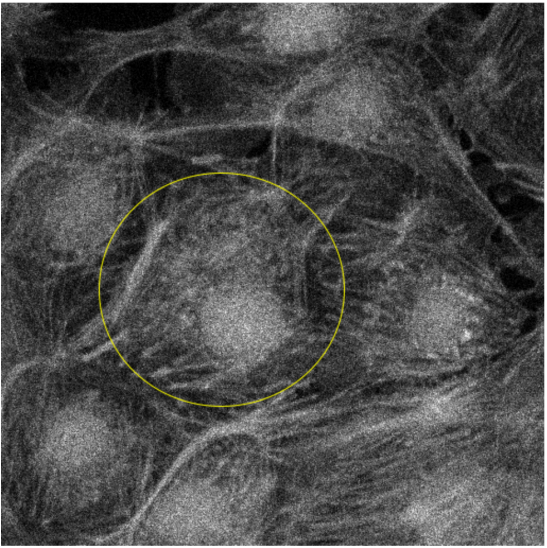

## C

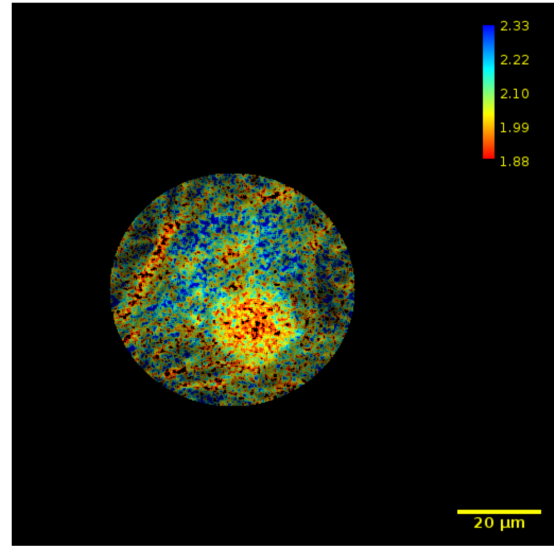

**Figure 8**: FLIM analysis shows a separation of three fluorescence components in a MitoTracker Red, Alexa Fluor 488 and DAPI labeled fixed BPAE cell sample (FluoCells Prepared Slide#1, ThermoFisher). **(A)** Excerpts from FLIM analysis script using SciJava Ops which was run in Fiji. **(B)** A 2D FLIM image of BPAE cells, imported into Fiji using Bio-Formats. An elliptical ROI has been drawn around a single cell using the ImageJ Elliptical selection tool. **(C)** FLIM analysis within SciJava Ops yields pseudo colored result images (with corresponding color bar), which could be further processed using other Ops. Note that script results were restricted to the elliptical ROI drawn in (A). The scale bar is 20 um and the FLIM lookup table reads 1.88ns to 2.33ns.

## 4.4 Spatially Adaptive Colocalization Analysis

Colocalization analysis remains a powerful tool to study the co-occurence of two signals of interest in a multi-channel dataset. Typically, workflows select a ROI, chart pixel data from both channels within the ROI on a scatter plot and determine the degree of colocalization by determining the colocalization quantification index (*e.g.* Pearson's, Li, *etc.*) (50,51). While effective, this strategy fails to account for the valuable spatial information in the rest of the image and relies upon a non-reproducible ROI. Wang *et al*. (52) address these shortcomings through the Spatially Adaptive Colocalization Analysis (SACA) algorithm. Unlike traditional colocalization approaches requiring a predetermined ROI the SACA algorithm determines colocalization strength at the pixel level (Figure 9C), considering all pixels within the image. By doing so SACA takes advantage of the spatial information within the image, avoiding ROI selection entirely.

Originally implemented in R, SACA has been translated into two Ops within the SciJava Ops Image library. The `coloc.saca.heatmapZScore` Op (Figure 9C) accepts two input images and returns the *z*-score heatmap of colocalization strength. Through incorporation within the Ops framework, the `coloc.saca.heatmapZScore` Op is able to depend upon `threshold.otsu` and `image.histogram` Ops, avoiding significant code duplication. The `stats.pnorm` Op accepts can be used with the z-score heatmap as an input to create a pixel wise *p*-value heatmap (Figure 9D). Similarly the `coloc.saca.sigMask` Ops works on the same *z*-score heatmap output from SACA and returns a binary mask indicating which pixels are significantly colocalized (Figure 9E).

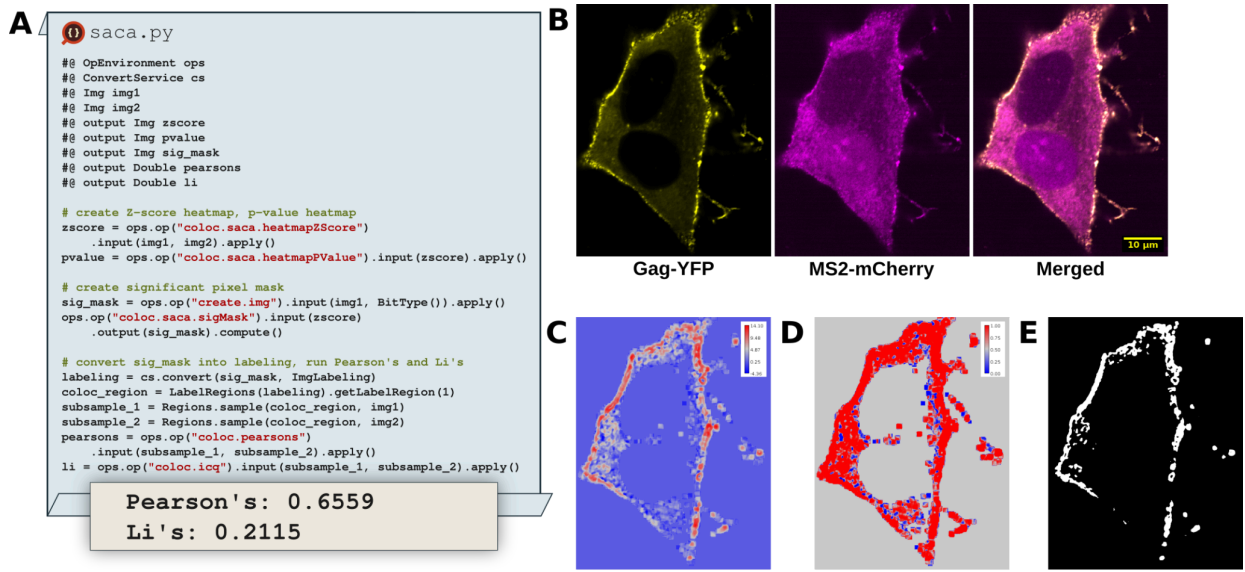

**Figure 9**: SACA on dual color fluorescent HIV-1 construct expressed in HeLa cells identifies colocalization regions of interest. **(A)** Excerpts of a Python Fiji script that utilizes the SACA framework implemented in SciJava Ops to detect regions of colocalization and determine the degree of colocalization. **(B)**

Input images of HIV-1 Gag-YFP, MS2-mCherry proteins, tracking colocalized viral particle production and viral mRNA trafficking dynamics at the plasma membrane. **(C)** *Z*-score heatmap of input data, indicating the strength of colocalization. **(D)** Pixel-wise p-value heatmap of the *Z*-score colocalization strength. **(E)** Significant pixel mask indicating which pixels are significantly colocalized in the input data.

## 4.5 Deconvolution

Although epifluorescence microscopy allows researchers to acquire sophisticated datasets, the images acquired still suffer from out-of-focus blur due to the light scatter through the sample. This problem can also be described as convolving the input image with the point spread function (PSF), the function that describes how light diffracts through the optical system. By performing the inverse of convolution (*i.e.* deconvolution) with the PSF, the out-of-focus blur can be removed, recovering the original image.

The SciJava Ops Image module implements both Richardson-Lucy (RL) and Richardson-Lucy Total Variation (RLTV) deconvolution (53), which improve image restoration quality by employing a regularization factor to suppress noise normally amplified by the RL algorithm (compare Figure 10B and 10C). Both implementations optionally allow vector acceleration (54) to speed up convergence and non-circulant boundary handling (55) to reduce artifacts near edges. SciJava Ops Image also implements the Gibson-Lanni algorithm, `create.kernelDiffraction`, to create a synthetic PSF (Figure 10D) with given imaging parameters including wavelength, and numerical aperture. Taken together, SciJava Ops allows researchers to perform 2D and 3D deconvolution with one framework.

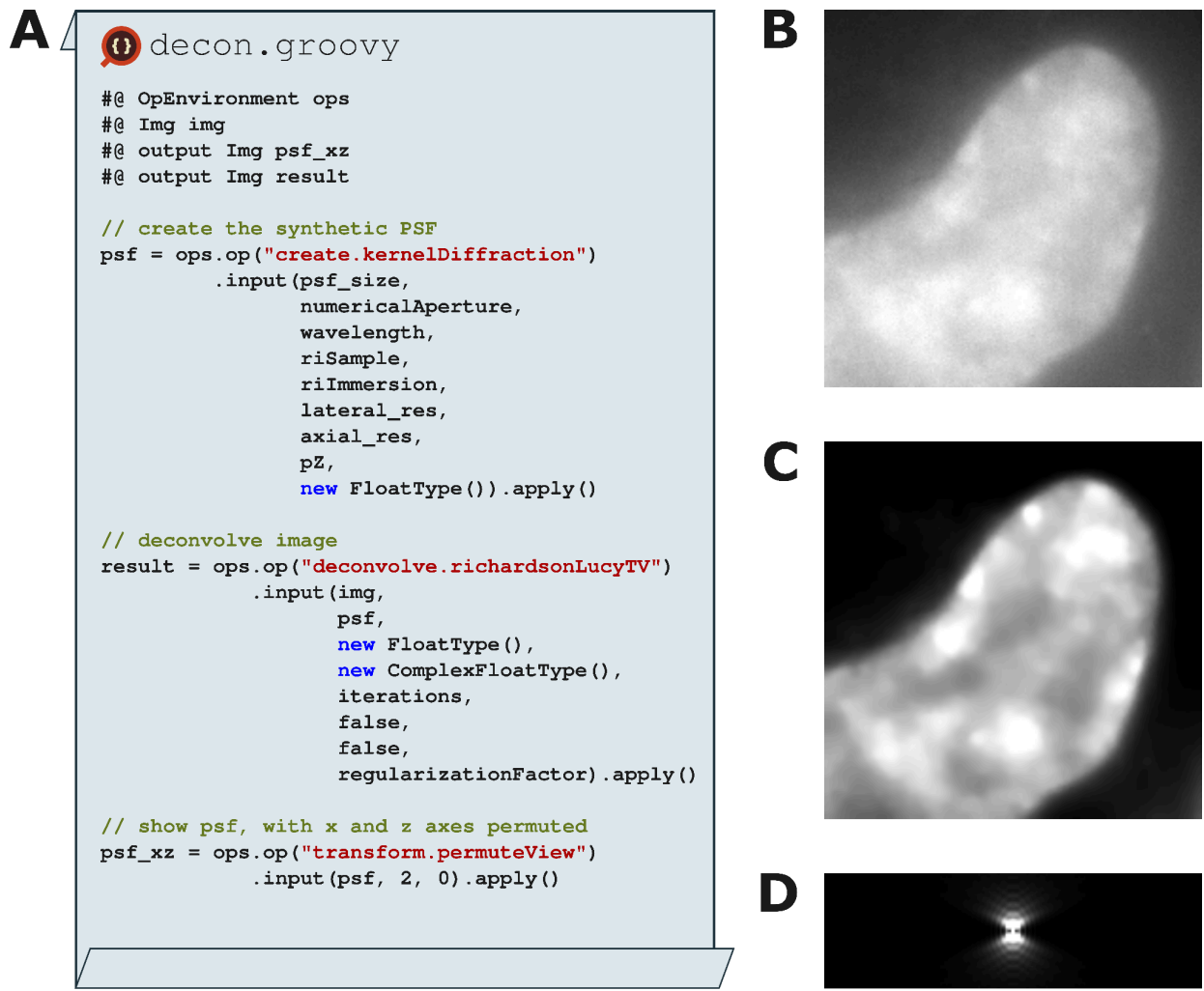

**Figure 10**: Synthetic point spread functions and Richardson-Lucy Total Variation deconvolution with SciJava Ops. **(A)** Excerpts of a Groovy Fiji script that utilizes the SciJava Ops Image module to create a synthetic point spread function (PSF) using experimentally derived parameters and perform Richardson-Lucy Total Variation (RLTV) on a 3D (X, Y, Z) dataset of a HeLa nucleus stained with DAPI. **(B)** Input raw data. **(C)** RLTV deconvolved result with 15 iterations. **(D)** XZ-project of the synthetic PSF generated using the Gibson-Lanni model.

## 5 Discussion

### 5.1 Scalability

When paired with the ImgLib2-cache framework, SciJava Ops can be used to generate or process images of virtually unlimited size. ImgLib2 supports very large, tiled images

(`LazyCellImg`) where tiles (Cell) are lazy-loaded when they are first accessed. A typical usage scenario is to load cells from disk on demand and keep recently used cells in a memory cache. However, cells can also be generated by arbitrary computations, cells can be modified and modifications written to disk when cells are evicted from cache, etc. Importantly, this all happens transparently, internal to ImgLib2, and is presented as a normal ImgLib2 image (`RandomAccessibleInterval`). This interacts with SciJava Ops in several ways:

First, such cached images can be used as input and output arguments for any Op on images. Many Ops can then simply work on larger-than-memory images, in particular Ops that use ImgLib2 data structures. Second, cells can be generated by Ops. Consider for example applying a per-pixel intensity correction to an input image. The result of this operation can be a `LazyCellImg` (with the actual computation only happening when pixels of this result are accessed). The cell loading functions would then run the intensity correction Op with the cell as the output image. Note, that in this case even Ops that require argument materialization can be applied, because they are run only on small portions of input/output image at a time. Finally, because the result of such an operation is just a (lazy evaluated) image, it can be used as input for another on-demand operation. We can pass the above intensity corrected image as the input argument to a cell loading function that uses a Gaussian convolution Op. The result is again a lazy image. When a pixel of this image is accessed, it will trigger the Gaussian cell loader to produce the cell containing the requested pixel. This will access pixels of its input, which will in turn trigger the intensity correction cell loader to produce the cell(s) containing those pixels.

## 5.2 Performance

A common pattern in software is to have a slow-but-general case, and a fast special case. Consider writing a method to create a new copy of an image. If we were to type its parameters on a general image data interface, we obtain the flexibility and extensibility to copy any sort of image; the actual data could be stored by column or by row, on GPU or in memory. If we were instead to restrict our method to images backed by primitive Java byte arrays, we could utilize that information to copy much faster (using `java.lang.System.arraycopy`). To provide the benefits of both choices, many programs employ case logic: if the image data is backed by an array in memory, copy it the "fast way", otherwise copy it the "slow but general" way element by element. Unfortunately, this logic not only makes programs more complex and difficult to maintain, but also prevents adding new cases without changing the source code. If we later want to add a case that copies image data between two buffers on a graphical processing unit (GPU) but we don't have access to the original source code, we must write a new method. In such situations, plugin-based frameworks like SciJava Ops allow developers to simply add another plugin to operate upon GPU buffers, and the matcher will take care of selecting this op in cases where the image data type indicates it would be compatible. This "extensible case logic" allows SciJava Ops to take advantage of future performance improvements with minimal maintenance.

While the SciJava Ops matcher allows future performance improvements, it is vital that that matcher must not impact performance itself. To quantify framework overhead, we defined a simple Op `benchmark.increment(data: byte[])`, defined as a static Java method, to increment the first byte in the provided array. This behavior was chosen to minimize the method's runtime so that runtime consists almost entirely of overhead. This Op was called both through direct invocation, and through the SciJava Ops framework. Each execution pathway was benchmarked using the Java Microbenchmark Harness (JMH) (https://github.com/openjdk/jmh) on a 2021 Dell Optiplex 5090 with Intel(R) Core(TM) i7-10700 CPU and 64 GB of DDR4 memory, and summarized with average execution times (second per invocation). Figure 11 visualizes the mean execution overhead of each profiled method, with error bars denoting the observed range. Direct (static method) invocation results show that our Op indeed imposes a negligible runtime of around 1 microsecond, ensuring that any time spent in other benchmarks are solely attributable to the SciJava Ops framework. Basic execution via SciJava Ops imposes an overhead of approximately 100 microseconds, however by using an internal cache SciJava Ops can reduce the overhead of repeated requests to approximately 3 microseconds. These results suggest that most of SciJava Ops' overhead comes from the matching process. As we would expect, the framework features of Op adaptation and parameter conversion provide additional overhead, and their composition shows that this overhead is additive. Perhaps most importantly, we show that the overhead imposed by SciJava Ops is significantly less than that of ImageJ Ops, with direct matching running approximately 70 times faster than its predecessor.

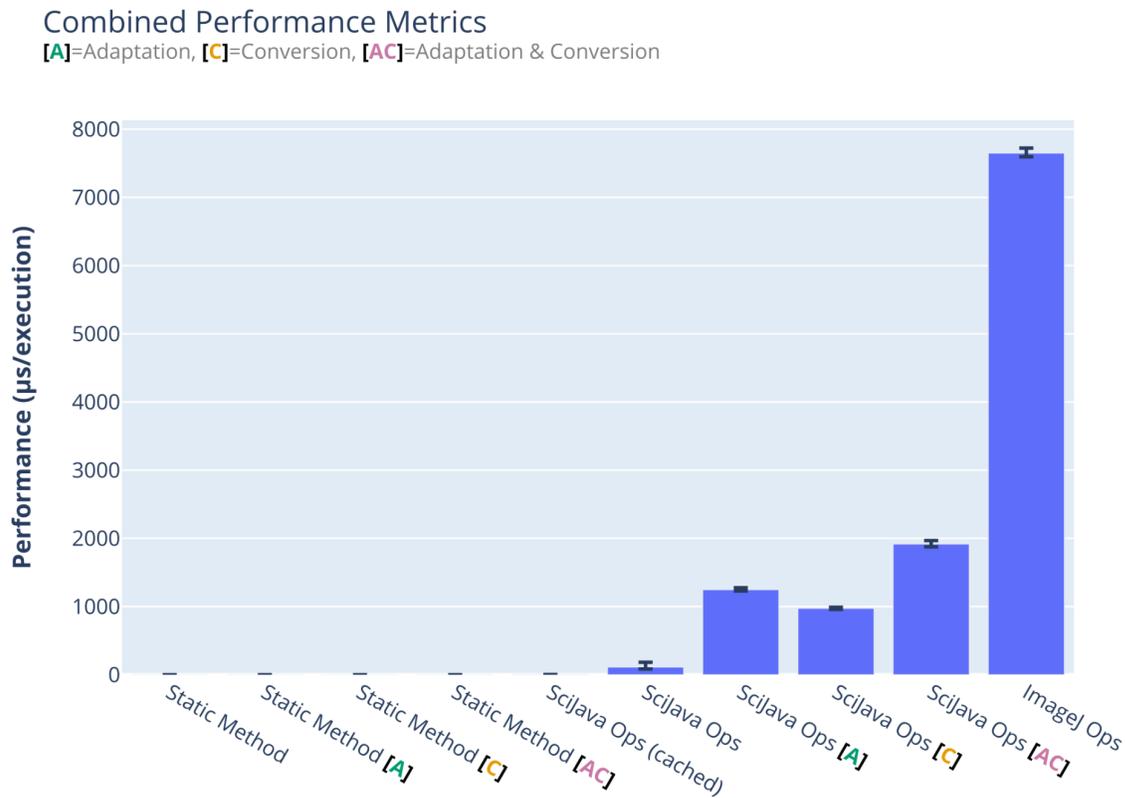

**Figure 11**: Execution overhead (in microseconds) of the Op `benchmark.increment(data: byte[])`, through *static method* invocation, *SciJava Ops*, and *ImageJ Ops*. Op adaptation [A] was additionally benchmarked for SciJava Ops by requesting

the Op as a function Op, and for static invocation by creating an output analogous to the Op adaptation routine. Parameter conversion [C] was benchmarked for SciJava Ops by requesting the Op with a `double[]` parameter, and conversion was performed manually for static invocation for comparison. Finally, a composition of adaptation and conversion [A+C] was benchmarked to profile the overhead of their combination. In all SciJava Ops benchmarks the Op cache was disabled to ensure the performance of the Op matcher was measured, except in the case where caching was explicitly enabled for the sake of comparison to static invocation.

## 5.3 Limitations

**For reproducibility, Ops must be deterministic.** Some algorithms are non-deterministic, with varying results upon multiple executions with the same inputs. This can happen when an algorithm uses a random number generator without a fixed seed value, or stores persistent data that affects the computation across executions, or uses multithreaded computation in a way that accumulates results differently depending on the timing of each thread's calculations. To introduce such algorithms in the Ops framework, they must be updated to avoid such non-deterministic behavior: use a fixed or user-provided seed for random number generation, accept all data as explicit inputs to the Op, and carefully coordinate all threads of a multithreaded computation to ensure reproducible calculation. This restriction ensures that Ops-based analyses produce the same results across computing environments, fostering reproducible science.

**Converting arguments between types may require copying data.** Parameter conversion incurs significant overhead whenever argument data must be copied. To that end, built-in engine.convert Ops wrap data whenever feasible, yet many conversions are infeasible without explicit data copying. For example, while ImgLib2 image structures allow zero-copy conversion of OpenCV `Mat` objects, the `Mat` interface provides no opportunity for zero-copy conversion in the other direction. Parameter conversion still allows SciJava Ops to provide seamless interoperability, but forced copies impact overall performance.

**The Op engine discriminates only on parameter types, not values.** With the current architecture of SciJava Ops, it is not possible to have two Ops with the same name, each operating on a different subset of the same input types. Consider the case of two-dimensional convolution; depending on kernel size, arithmetic convolution may or may not be faster than addition within the frequency domain. Since a kernel's size cannot be inferred from its type, the SciJava Ops framework cannot delegate between these two algorithms. Instead, delegation would require a single Op to wrap both algorithms with conditional logic, limiting extensibility.

**Op executions are complicated to trace.** While the SciJava Ops Framework is built to foster reproducible workflows, it can be difficult to determine which blocks of code correspond to each workflow component. Some ambiguity comes from the declarative nature of Op requests, which necessarily adds a layer of indirection. Furthermore, Op adaptation, parameter conversion and other internal engine features further obfuscate the underlying Op through additional transformations. To obtain an understanding of which Ops were run, two different components of the SciJava Ops framework can be used. Each Op contains an InfoTree data structure, which provides a full graph of dependencies, adaptations and parameter conversions. Additionally, the `OpHistory` keeps a record of each Op that edited an output buffer, enabling reconstruction of the set of Ops used to produce a given Object. While we do not predict SciJava Ops users to need access to these features often, it is nonetheless an extra level of complication compared to direct imperative function calls.

**The Ops engine runs on the Java Virtual Machine.** The JVM is a powerful piece of technology, offering good performance and a well-designed type system. But the OpenJDK platform providing the JVM can weigh in at hundreds of megabytes in size, and may be undesirable as a dependency in scenarios with limited space such as embedded systems. While the architecture of SciJava Ops is not fundamentally tied to Java or the JVM specifically, the current implementation is, and more effort would be necessary to change that—e.g. implementing another Ops engine on top of a different platform, or adjusting the Ops codebase to work with an Ahead-of-Time compiler such as Kotlin Native (https://kotlinlang.org/docs/native-overview.html) or GraalVM's Native Image (https://www.graalvm.org/jdk21/reference-manual/native-image/).

**Naming is hard.** As the library of available Ops grows, care must be taken to ensure consistent naming across different collections of Ops. While there is no requirement that two Ops with the same name produce binary-identical results when passed the same inputs—and indeed such a requirement would be very challenging to achieve in practice—it is a judgment call whether two pieces of code performing variations of the same idea are close enough to be considered the "same algorithm" and therefore be labeled with the same Op name. Like any ontology project, guidelines will need to emerge and collections will need to be well curated in order to prevent the growth of an inconsistent mess over time.

## 5.4 Future Directions

The current state of SciJava Ops provides the core frameworks for extensibility, interoperability and reproducibility, with a number of standard image processing algorithms from SciJava Ops Image, ImgLib2, and OpenCV bindings. This is all in service of progress towards a single unified cross-platform algorithm entry point. We believe the next steps in this path are as follows.

### 5.4.1 Wrapping new libraries from Python, C++, and other software languages

There are many existing libraries that, when made usable from SciJava Ops, will further expand workflow possibilities. Many gaps in SciJava Ops Image can be filled using libraries built upon the Python scientific stack, such as SciPy, scikit-image, scikit-learn, and PyTorch. C++ libraries built atop CUDA and GPUs can be wrapped to increase performance over existing algorithms. The result will be full access to the most popular algorithms from these communities, all using the same syntactically consistent Op builder pattern, facilitating more seamless combination of algorithms from a diversity of sources.

### 5.4.2 Integrate SciJava Ops into other scientific applications

In this text we have highlighted Fiji as a natural entry point for SciJava Ops, but as we have shown SciJava Ops can be used in environments outside of Fiji as well. Beyond the current Python support powered by scyjava, users can also write standalone scripts from any Java-based scripting language (e.g. Groovy, Jython or Kotlin). We plan to continue working with the imaging software community to integrate SciJava Ops into additional image analysis tools (e.g. napari, CellProfiler, Icy) so that the same algorithms can be used the same way across applications, which will not only bring value to the users of these tools, but also encourage the future integration of additional algorithms libraries into Ops as an easier route to dissemination.

### 5.4.3 An algorithms index for algorithm discovery and exposure

Today there is already a question of "what image processing algorithms are available?" and simply expanding SciJava Ops to integrate more algorithm libraries will not answer this question by itself. We envision a unified web-based algorithm repository, with clear routes to identifying which algorithms are available and which library(ies) implement said algorithm, to maximize accessibility. Developers can browse such an index to determine whether their algorithm has already been implemented in another project and to calculate the comparative burden of wrapping an existing algorithm instead of reimplementing it within their own library. They can also publish to such an index to expose their new implementations, avoiding reimplementation done purely because another developer could not find any existing code. Users can browse such an index to determine which library might best suit their needs, rather than extensively searching through the documentation of many different packages.

We believe that SciJava Ops can ultimately be leveraged to build such an algorithm repository, providing users and developers with both the means to locate algorithms of interest and also practical instructions for including them in the workflow environments they prefer. The index of algorithm implementations will be automatically generated and maintained from an aggregated collection of known Ops, while also featuring a general page about each algorithm (e.g. filter.gauss) with a community-curated wiki-style introductions to the algorithm along with links to all known implementations across various software packages.

# 6 Conclusion

SciJava Ops represents the culmination of our experience creating extensible frameworks for use in scientific communities. Designed as a foundational layer for algorithms across the scientific image analysis community, it aims to meet technical requirements across many segments of a broad target audience: strong typing guarantees for developers, declarative calls for users, extensible code-free integration paths coupled with easy-to-use annotations, with automatic adaptation and type conversion to aid longevity and maintainability of any algorithm library in our ecosystem. We believe we have succeeded at creating a framework capable of bridging the gaps between libraries of algorithms in Java and beyond, facilitating reproducible science with less time lost to recapitulation and fragmentation. We are excited for the potential for growth and contributions as SciJava Ops enters into public use and continues to grow, approaching our vision of a consolidated image processing ecosystem, where users can discover and harness algorithms across many software technologies in a unified way.

# 7 Methods and Materials

## 7.1 RLTV deconvolution and Python (Scyjava) datasets

HeLa and 3T3 cells were plated in 8-well no. 1.5H glass bottom slides (Ibidi) and maintained at 37°C, ~50% humidity, and 5% $CO_2$ prior to fixation in 4% paraformaldehyde in PBS. Both HeLa and 3T3 fixed cells were stained with 4′,6-diamidino-2-phenylindole (DAPI). Z-stack images with a step size of 0.1 μm were acquired on a Nikon Ti-Eclipse inverted wide-field epifluorescent microscope (Nikon Corporation) using a 1.45NA 100x Plan Apo oil immersion objective lens and an ORCA-Flash4.0 CMOS camera (Hamamatsu Photonics, Skokie, IL USA). DAPI images were acquired with a 325-375 nm emission / 435-485 nm excitation filter set.

## 7.2 SACA dataset

HeLa cells were in 24-well glass bottom plates (Cellvis) and fixed in 4% paraformaldehyde and stained with DAPI 48 hours post transfection with a dual color fluorescent $HIV_{NL4-3}$ construct expressing Gag-YFP and MS2-mCherry fusion proteins. Images were acquired on an A1R laser scanning confocal microscope built on Nikon Ti-Eclipse inverted base using a 1.4NA 60x Plan Apo oil immersion objective lens. DAPI, YFP and mCherry were excited by the 405 nm, 488 nm and 561 nm diode lasers respectively.

## 7.3 FLIM dataset and analysis

Fixed stained slides of bovine pulmonary artery endothelial cells (BPAEC) were obtained from ThermoFisher Scientific (FluoCells™ Prepared Slide #1, CAT#F36924). The sample contains MitoTracker™ Red CMXRos, Alexa Fluor™ 488 Phalloidin, and DAPI nuclear stains and further details can be obtained from the manufacturer. The sample was imaged using a 40xWI 1.15NA Nikon objective lens with multiphoton excitation at 740nm and collected with a 650nm shortpass filter to image all three fluorophores. The imaging and scanning was controlled by Openscan-LSM (https://github.com/openscan-lsm/OpenScan) and SPC180 (Becker-Hickl Gmbh, Berlin, Germany) fast timing electronics card. The FLIM data was modeled using a mono-exponential decay and Levenburg Marquardt algorithm (LMA) fitting using SciJava Ops FLIM.

## 8 Project Information and Data Availability

- **Source code:** https://github.com/scijava/scijava
- **Documentation:** https://ops.scijava.org/
- **Data Availability:** https://ops.scijava.org/en/1.0.0/paper-2024
- **Operating system:** Linux, macOS, Windows
- **Programming language:** Java SE 11
- **License:** Simplified BSD License

## 9 Conflict of Interest

The authors declare that the research was conducted in the absence of any commercial or financial relationships that could be construed as a potential conflict of interest.

## 10 Author Contributions

CTR, CB, and JS designed and implemented the ImageJ Ops framework. CB, CTR, TOB, SH, BN, AW, and GJS implemented analysis routines for Ops. GJS, DK, MW, MCH, and CTR developed SciJava Ops, revised from the original design of ImageJ Ops. MCH reviewed the code, fixed bugs, and integrated OpenCV. ELE exercised the Ops framework with use cases and expanded the project's documentation. TP and SS provided feedback on the Ops design and conceptualized the ImgLib2-cache-based Ops mechanism for greater scalability. MRB and KWE supervised and supported development and organized collaborative events to drive progress. All authors contributed to the manuscript text.

## 11 Funding

We acknowledge funding support from the Chan Zuckerberg Initiative (CTR and KWE), NIH Grants P41GM135019 (KWE) and U54CA268069 (KWE) as well as additional support from the Morgridge Institute for Research (ELE and KWE).

## 12 Acknowledgments

The authors acknowledge Barry DeZonia for his initial effort to develop a functional image processing library for ImgLib2; Martin Horn for co-founding the ImageJ Ops project; Stephan Preibisch for co-developing the ImgLib2 library; Deborah Schmidt for helpful ideas and discussions; Jenu Chacko for FLIM data and analysis guidance; Dasong Gao for his work on the FLIMJ Ops; Sydney Lesko for confocal data for SACA; and all the software developers who contributed to Ops over the years.